\documentclass[journal=jpccck,manuscript=article, layout=twocolumn]{achemso}

\usepackage{achemso}
\usepackage{amsmath}
\usepackage{graphicx}
\usepackage{dcolumn}
\usepackage{bm}
\usepackage{color}
\usepackage{hyperref}

\title{Inverse Design of a Graphene-Based Quantum Transducer via Neuroevolution}

\author{Kevin Ryczko}
 \affiliation{Department of Physics, University of Ottawa, Ottawa, ON K1N 6N5, Canada}
 \email{kevin.ryczko@uottawa.ca}
 \alsoaffiliation{Vector Institute for Artificial Intelligence, Toronto, OM M5G 1M1, Canada}

\author{Pierre Darancet}
\email{pdarancet@anl.gov}
\affiliation{Center for Nanoscale Materials, Argonne National Laboratory, Lemont, IL 60439, United States}
\alsoaffiliation{Northwestern Argonne Institute of Science and Engineering, Evanston, IL 60208, United States}

\author{Isaac Tamblyn}
\email{isaac.tamblyn@nrc.ca}
\affiliation{National Research Council of Canada, Ottawa, ON K1N 5A2, Canada}
 \alsoaffiliation{Department of Physics, University of Ottawa, Ottawa, ON K1N 6N5, Canada} \alsoaffiliation{Vector Institute for Artificial Intelligence, Toronto, OM M5G 1M1, Canada}
\date{\today}
\begin{document}

\begin{abstract}
We introduce an inverse design framework based on artificial neural networks, genetic algorithms, and tight-binding calculations, capable to optimize the very large configuration space of nanoelectronic devices. 
Our non-linear optimization procedure operates on trial Hamiltonians through superoperators controlling growth policies of regions of distinct doping.
We demonstrate that our algorithm optimizes the doping of graphene-based three-terminal devices for valleytronics applications, monotonously converging to synthesizable devices with high merit functions in a few thousand evaluations (out of $\simeq 2^{3800}$ possible configurations).
The best-performing device allowed for a terminal-specific separation of valley currents with $\simeq 96$\% ($\simeq 94\%)$ $K$ ($K'$) valley purity. 
Importantly, the devices found through our non-linear optimization procedure have both higher merit function and higher robustness to defects than the ones obtained through geometry optimization. 
\end{abstract}

\maketitle

\section{Introduction}
Tailoring the properties of nanoelectronics devices leveraging quantum phenomena without classical analog is central to spintronics~\cite{bader2010spintronics}, valleytronics~\cite{schaibley2016valleytronics}, quantum transduction, quantum sensing, and quantum computing \cite{Mak2012ControlHelicity,Zeng2012ValleyPumping,Cao2012Valley-selectiveDisulphide,Mak2014TheTransistors}. 
The figure of merit of such nanoelectronics devices strongly depends on the details of the low-energy Hamiltonian and can be impacted by numerous energy scales. For example, in the case of graphene-based devices~\cite{CastroNeto2009TheGraphene},  valley-polarized currents can be manipulated spatially by tuning the edges of the devices \cite{Rycerz2007ValleyGraphene, xiao2007valley}, through strain-engineering~\cite{stegmann2018current, jones2017quantized, levy2010strain, guinea2008gauge, settnes2016pseudomagnetic, settnes2016graphene, zhai2018local, stegmann2016current, wu2017quantum, andrade2019valley, hatsugai2019so, mcrae2019graphene, fujita2010valley},  defect-engineering~\cite{gunlycke2011graphene, chen2014controlled}, and by the combination of edge termination and doping via a gate voltage or molecular adsorption ~\cite{Rycerz2007ValleyGraphene, Cheianov2007TheJunctions, park2019magnetoelectrically, garcia2008fully, aktor2019topological, natan2007electrostatic}. While these studies predict high valley-filtering (e.g. with valley purity exceeding  $\sim$ 95\%~\cite{Rycerz2007ValleyGraphene}), actual, synthesizable, devices will display a superposition of these effects, resulting in a complex optimization process of these possibly-competing energy scales. 

Such optimization is an inverse design problem, where one knows the desired output of the device, as defined by a simple merit function (e.g. valley or spin purity, current magnitude, etc.), but does not know the optimal way to modify the device under experimental and synthesizability constraints to obtain local or global extrema of these merit functions. Due to their large tunability and the number of atoms they contain (typically exceeding $10^{3}$), the configuration space of nanoelectronic devices is extremely large and impossible to fully explore with high-throughput forward quantum transport solvers \cite{Groth2014Kwant:Transport, steiger2011nemo5}. In this context, the coming-of-age of artificial intelligence approaches capable of optimizing in large configuration space \cite{silver2017mastering} offers significant opportunities. Local optimization techniques were used, for example, to design physically-unintuitive demultiplexer devices~\cite{Piggott2015InverseDemultiplexer, doi:10.1021/acsphotonics.7b00987} and chemical compounds~\cite{LeardiGeneticReview}. These approaches could also be accelerated by machine learning the quantities of interest, as demonstrated in Ref. \cite{torres2019valley}.

Efficiently searching through large configuration spaces generally implies sampling a lower-dimensional space that maps to the original, higher dimensional, search space, possibly constructing this mapping with machine learning \cite{melati2019mapping}. Another possibility is to use simple policies that yield complex outcomes, mimicking non-linear optimization processes observed in nature. In the work of Schelling \cite{Schelling1971DynamicSegregation}, the complex dynamics of segregation was modeled by subtle changes in the underlying policies. This was also shown to be the case by Wolfram \cite{wolfram1983statistical} on cellular automata. The search of a large configuration space that is represented by an artificial neural network (ANN) can be done with GAs or gradient descent. GAs are gradient-free, are less likely to be trapped in local minima, and converge rapidly. On the contrary one must have a flexible ansatz (or set of genes) to fully describe the configuration space, and it is difficult to include a feasibility of fabrication score in the objective function. This could result in finding optimal structures with a GA that could not be fabricated in the laboratory. This issue is eliminated for generative machine learning models where gradients are used to minimize an objective function. Machine learning models are flexible with respect to input and ability to learn complex mappings, but one must initially train the model and be confident in the model's ability to perform inference over a wide range of structures.

In this work, we demonstrate inverse design of nanoelectronic devices using growth policies coupled to ANNs, GAs, and tight-binding calculations. The choice of GA was made because there is no analytic connection between the objective function and the policy network. Our framework is capable to navigate the very large configuration space through a non-linear optimization procedure that operates on trial Hamiltonians through a superoperator, effectively controlling growth policies of regions of distinct doping.
We demonstrate optimization of the doping profile of graphene-based three-terminal devices for valleytronics applications, monotonously converging to synthesizable devices with high merit functions in a few thousand evaluations (out of $\simeq 2^{3800}$ possible configurations).
The best-performing device allowed for a terminal-specific separation of valley currents with $\simeq 96$\% ($\simeq 94\%)$ $K$ ($K'$) valley purity.
Importantly, the devices found through our non-linear optimization procedure have both higher merit function and higher robustness to defects than the ones obtained through linear geometrical optimization \cite{SI}.

In the Methods section, we detail our theoretical approach which includes the structure design, optimization procedure, and the methodology for calculating valley polarized currents. In the Results section, we show that our optimization procedure produces structures that can separate valley polarized currents and analyze the GA optimization procedures.

\section{Methods}
\label{methods}

\begin{figure}[H]
    \centering
    \includegraphics[width=0.90\linewidth]{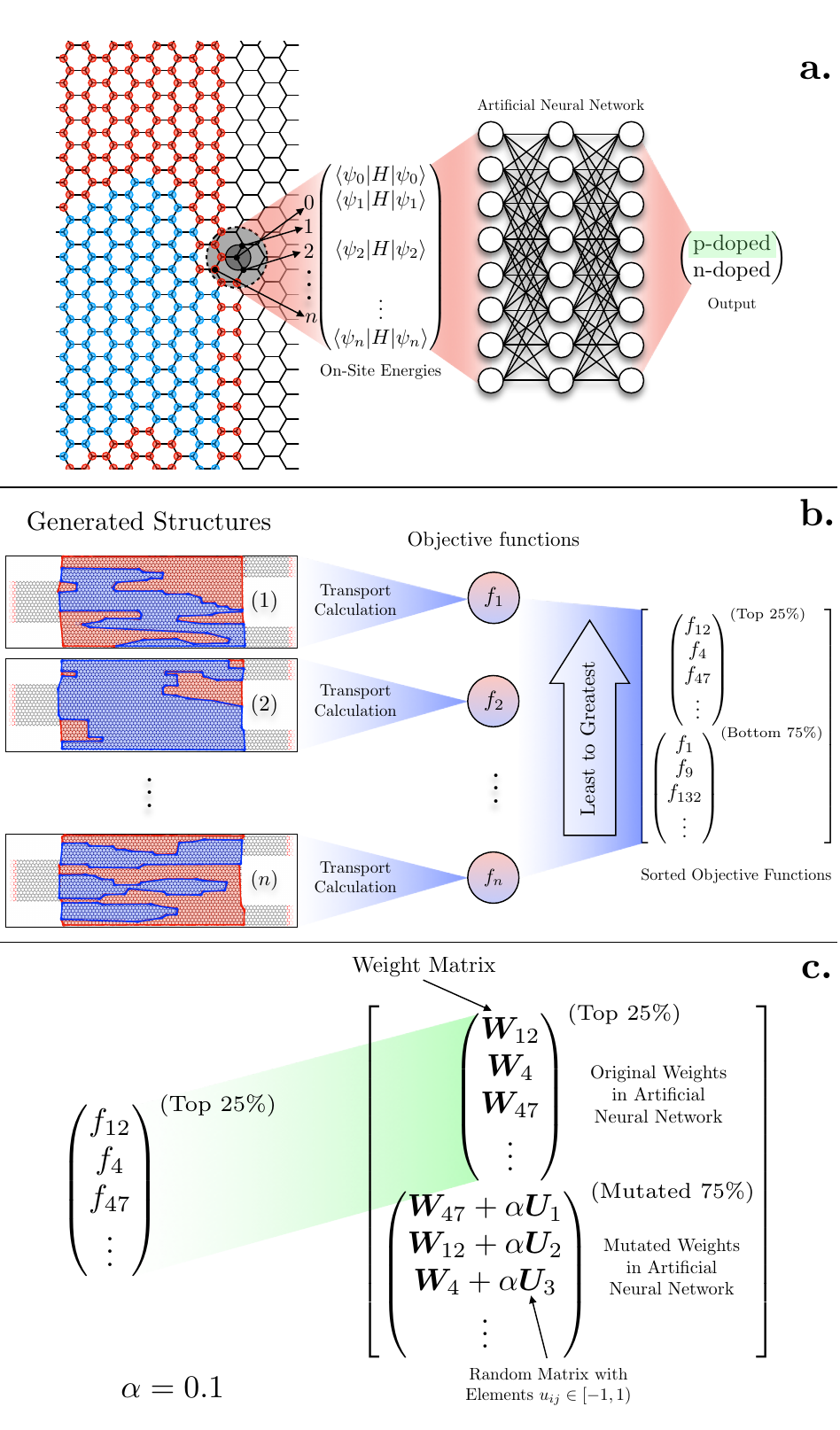}
    \caption{\label{device_design} Schematic description of optimization procedure. Starting from a trial Hamiltonian (consisting of either p- or n-doped regions on a fixed device lattice) (a), an artificial neural network policy is used to modify the on-site energy based on the on-site energies of its neighboring sites (nearest and next-nearest neighbors). The on-site energies for this set of lattice sites are concatenated into an input vector which is fed into an ANN policy that determines whether the selected lattice site will be p- or n-doped. The resulting merit function of each trial structure is then computed with a tight-binding transport solver~\cite{Groth2014Kwant:Transport} (b). The structures shown were taken from a random optimization procedure, with regions of n- and p-doping highlighted for clarification. The structures are then ranked based on their objective functions. The structures in the next generation are produced (c) using a genetic algorithm on the policies of the top 25\% of the performers in the population of previous generations.}
\end{figure}

Our general workflow is summarized in Fig. \ref{device_design}, and consists of a structure generation scheme using an ANN, a (forward) quantum transport solver, and a GA.

Our framework operates on a fixed device structure, as represented by a graph Hamiltonian with a fixed adjacency matrix. Each vertex of the graph represents an orbital contributing to the low-energy Hamiltonian. 

Each vertex has a corresponding on-site energy and a set of nearest neighbour couplings with sites defined in the adjacency matrix. The Hamiltonian of the system is 
\begin{equation}
\label{tight_binding_hamiltonian}
H = \sum_{\langle i, j\rangle} \tau_{ij}|i\rangle \langle j| + \sum_{i}U_i | i\rangle \langle i |,
\end{equation}
where the first summation of Equation \ref{tight_binding_hamiltonian} describes the nearest neighbour interactions between orbitals $i$ and $j$, and the second summation describes the on-site potential. 
The device is contacted to ballistic terminals with a predefined set of undoped lattice sites. 

We focus on the case of three-terminal devices  comprised of 3846 sites on a model of doped graphene, using
 $\tau_{ij}=-2.7$ eV as done in \cite{CastroNeto2009TheGraphene}, and
\begin{eqnarray}
U_i = \left\{
\begin{array}{cc}
U & \text{ n-doped}\\
-U & \text{ p-doped} \\
0 & \text{terminal}
\end{array}
\right. 
\end{eqnarray}
with $U=0.2$ eV, in alignment with past work \cite{wakabayashi2002electrical}. This corresponds to a device with a body whose length is 128~\AA. We also studied devices with shorter body lengths (down to 64 \AA) and found no quantitative change in our results.

The spatial extent of the p- and n-doped doped regions on a given device is the result of an ANN policy that outputs the probability of given lattice sites to be p- or n-doped, as we now describe. 

At the first iteration, we set all of the on-site energies to $U_i=0.5$ eV. The choice of setting the on-site energies initially to 0.5 eV is arbitrary and the initialization can bias the optimization. We found that an initialization value $< 0.5$ biases $K$ current and $> 0.75$ biases $K'$ current. This bias is small for values close to the transition point but becomes non-negligible the further one moves from the transition point. For each lattice site $i$, the on-site energy along with the on-site energies of its nearest and next-nearest neighbors are concatenated into an input vector ${\bf v}_{\text{input}}$. This choice of input vector is motivated by the structure of the nearest-neighbor tight-binding Hamiltonian.

This choice of input vector is equivalent to using a graph convolution layer \cite{NIPS2015_5954, kipf2016semi} where the weight matrix is shared amongst all atoms. The input vector ${\bf v}_{\text{input}}$ is then fed into the ANN policy where the output of the ANN is a number $p\in[0,1]$. A random number $u_p\in[0,1)$ is then drawn such that if $p < u_p$ the site is p-doped, and n-doped otherwise. This ANN policy can also be thought of as a superoperator that operates on a constant tight-binding Hamiltonian yielding a Hamiltonian with a set of desired optimized properties. At each iteration, the weights of the ANN policy are updated using a GA in conjunction with the ANN~\cite{beeler2019optimizing, such2017deep}. 

To avoid producing structures where the doping changes over a short length scale, we apply Gaussian blurring as well as binary erosion and dilation on binary images of the lattice (1(0) represented p(n)-doping). Binary images were created by first initializing an array of zeros with dimensions equal to the length (128 \AA) and width (64 \AA) of the body. This made for an image with size $128\times64$. Afterwards, pixels were set to 1 if p-doped carbon atoms fall into a respective pixel. A Gaussian blur was then applied with a standard deviation of 2 \AA, followed by a threshold operation with a value of 0.5, a binary erosion operation, and a binary dilation with a 50\% chance. All of these operations can be written as kernels that operate, as a convolution, on the binary image. In addition, they can all be found in the scikit-image Python package \cite{van2014scikit}. We note that this post-processing protocol allowed for a 9\% improvement in the objective function.

\begin{figure*}[H]
	\centering
    \includegraphics[width=0.98\linewidth]{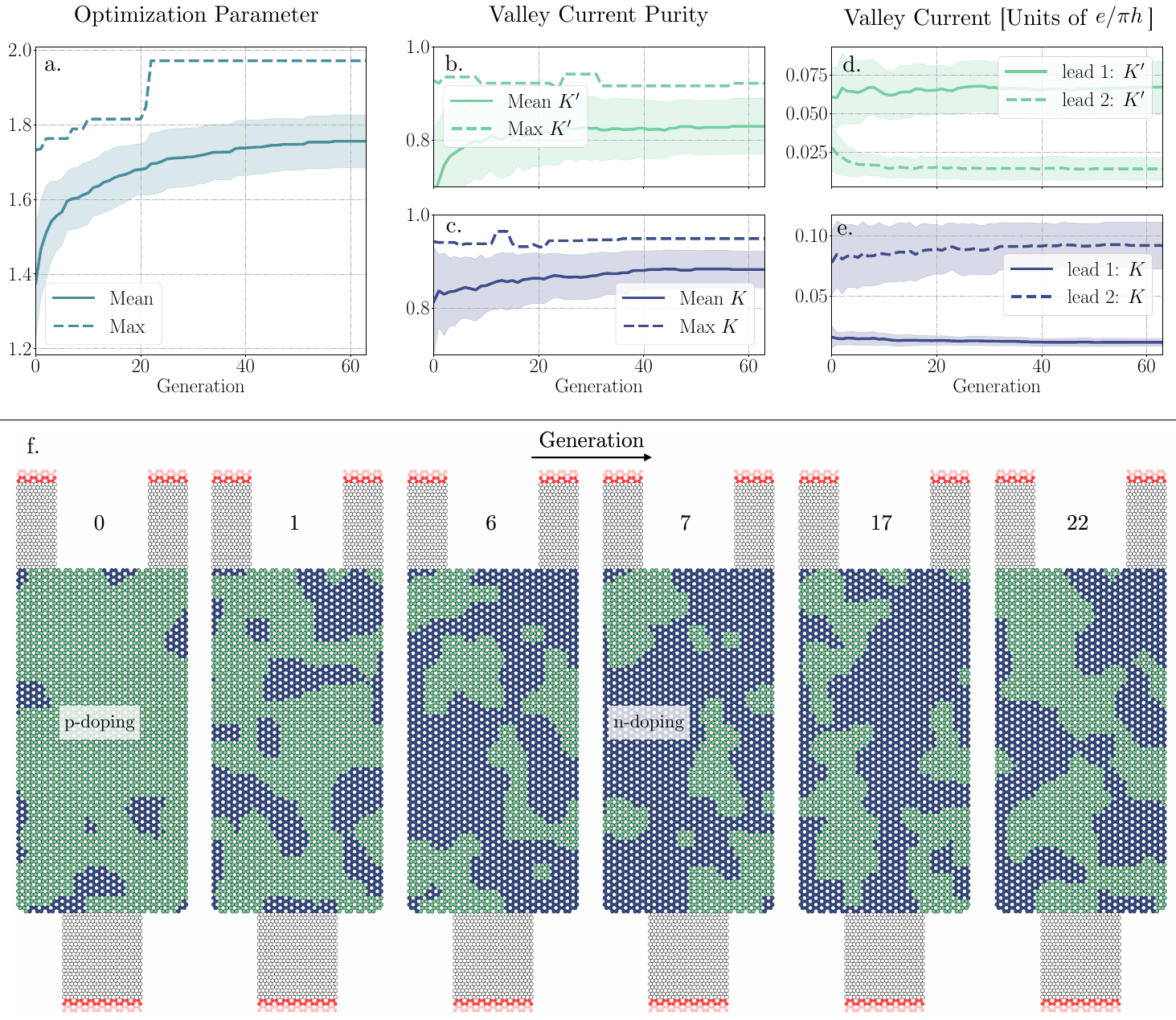}
    \caption{ Optimization function (a), valley current purity (b-c), and valley currents (d-g) recorded during the optimization processes using the artificial neural network policy. For each independent optimization, the devices with the maximum objective functions are recorded. The solid lines show the average values of these devices and the lighter regions show the standard deviations. In f. we show the evolution of best performing structures for one seed.}
    \label{fig:results_1}

\end{figure*}

The initialization stage of the optimization process consists of $N=160$ devices, each corresponding to a distinct ANN policy. 

 Given these $N$ Hamiltonians, we calculate the valley-polarized currents using the Kwant Python package \cite{Groth2014Kwant:Transport}, through the following equation:
\begin{equation}
I_{K/K'} = \frac{e}{\pi h}\int_{-\infty}^{\infty} d\epsilon~G_{K/K'} [f(\epsilon; \mu_L) - f(\epsilon ;\mu_R)]
\label{eq:ValleyCurrent}
\end{equation}
where $G_{K/K'}$ is the valley-dependent conductance, $f(E; \mu_{L/R})$ is the Fermi-Dirac function for the left ($L$) and right ($R$) leads, $e$ is the electron charge, and $h$ is Planck's constant.  To calculate the valley-polarized conductance we use the Landauer-B\"uttiker formula
\begin{equation}
G_{K/K'} = \frac{2e^2}{h}\sum_{n_{\text{modes}, K/K'}}T_n,~~~~~T_n=\sum_{m_{\text{modes}}}|t_{nm}|^2
\end{equation}
where the sum over $n_{\text{modes}, K/K'}$ are for incoming modes with momenta associated with valley $K$ or $K'$, and the sum over $m_{\text{modes}}$ are for all out going modes in the lead where we want to measure the current. Travelling modes (either $K$ or $K'$) are identified by considering their velocity and momentum, following~\cite{Rycerz2007ValleyGraphene}. The transmission probabilities $T_n$ are computed after the determination of the scattering matrix. In addition, we use a bias of 0.5 eV and a grid spacing of 1 meV to evaluate the integral in Eq.~\ref{eq:ValleyCurrent} throughout all of our calculations.

We consider three-terminal devices with a non-valley-polarized current incoming from lead $0$ (the ``left'' of the device in Fig. \ref{device_design} b). At the opposite side of the device, two leads $1$ and $2$ at identical chemical potentials collect the valley-dependent current injected from lead $0$. 

Before defining the objective function, we first define the purity of $K'$ current in lead 1 to be $P_{K',1} = I_{K', 1} / (I_{K, 1} + I_{K', 1})$ and the purity of $K$ current in lead 2 to be $P_{K, 2} = I_{K, 2} / (I_{K, 2}+ I_{K', 2})$. We then define the normalized total current $I_{\text{total}} = (I_{K', 1} + I_{K, 2}) / I_P $ where $I_P=0.3~e/\pi h$ is the total current of the pristine graphene lattice. We search for structures that maximize the multivariate objective function:
\begin{eqnarray}
\label{optimization_function}
    F(I_{K, 1}, I_{K', 1}, I_{K, 2}, I_{K', 2}) &=& P_{K', 1}^2 + P_{K, 2}^2 + I_{\text{total}}^2.\nonumber\\
\end{eqnarray}
 
We compute Eq. \ref{optimization_function} for each of the $N$ devices. Only the ANN policies associated with the top 25\% devices are kept to populate the next generation, through random mutation of the weights of the ANN policy yielding a new device. We do not consider crossover mutations in our study.

The weights of the newly generated ANN policy are
      \begin{equation}
         \begin{pmatrix}
         w_1\\
         w_2\\
         \vdots\\
         w_M
         \end{pmatrix}_{\text{new}}=\begin{pmatrix}
         w_1\\
         w_2\\
         \vdots\\
         w_M
         \end{pmatrix}_{\text{old}}+\alpha\begin{pmatrix}
         u_{1}\\
         u_{2}\\
         \vdots\\
         u_{M}
         \end{pmatrix}
     \end{equation}
where $\alpha=0.1$ is a small parameter (similar to a learning rate), $u_i\in[-1, 1]$ is a random number, and $M$ is the total number of weights. In our case we used ANNs with 2 hidden layers each with 128 neurons. The logistic activation function was used throughout the ANN. During the optimization process we ran calculations for 160 devices in parallel and performed 64 generations. The calculations performed for each generation took $\simeq45$ minutes on single node with 40 CPUs.

\section{Results}
\label{results}

The results of the optimization procedure are shown in Figure \ref{fig:results_1}. We show the average and maximum optimization function, valley current purity, and associated valley currents for the best performing devices of each independent seed as a function of generation following the Methods section. We also show the associated standard deviation over the $N=64$ independent calculations. Both the averages of the objective function and purity of the valley currents converge monotonously across 64 generations. The optimization process produced a structure with maximum purity of 96\% (94\%) $K$ ($K'$) valley current. This result is comparable to Rycerz \textit{et al.} \cite{Rycerz2007ValleyGraphene} where they achieved $\sim95\%$ purity with an idealized valley filtering device.

The average purity of $K$ valley current at generation 0, based on the random initialization is higher ($>80\%)$ than the average purity of $K'$ valley current for the best performing devices. In the remainder of the optimization process, the algorithm is capable to maintain the $K$ valley current purity steady while increasing the value of the $K'$ valley current purity, which started at an average value of $\sim70\%$ purity.

In addition to the convergence of the average currents, average measures of purity, and the average of the objective functions we also find that the standard deviation of the optimizations decreases as a function of generation. This indicates that the population is converging to a similar local maxima. 

In Figure \ref{fig:results_1} f, we also show the evolution of the best-performing structures for a single seed.

\begin{figure}[H]
    \centering
    \includegraphics[width=0.99\linewidth]{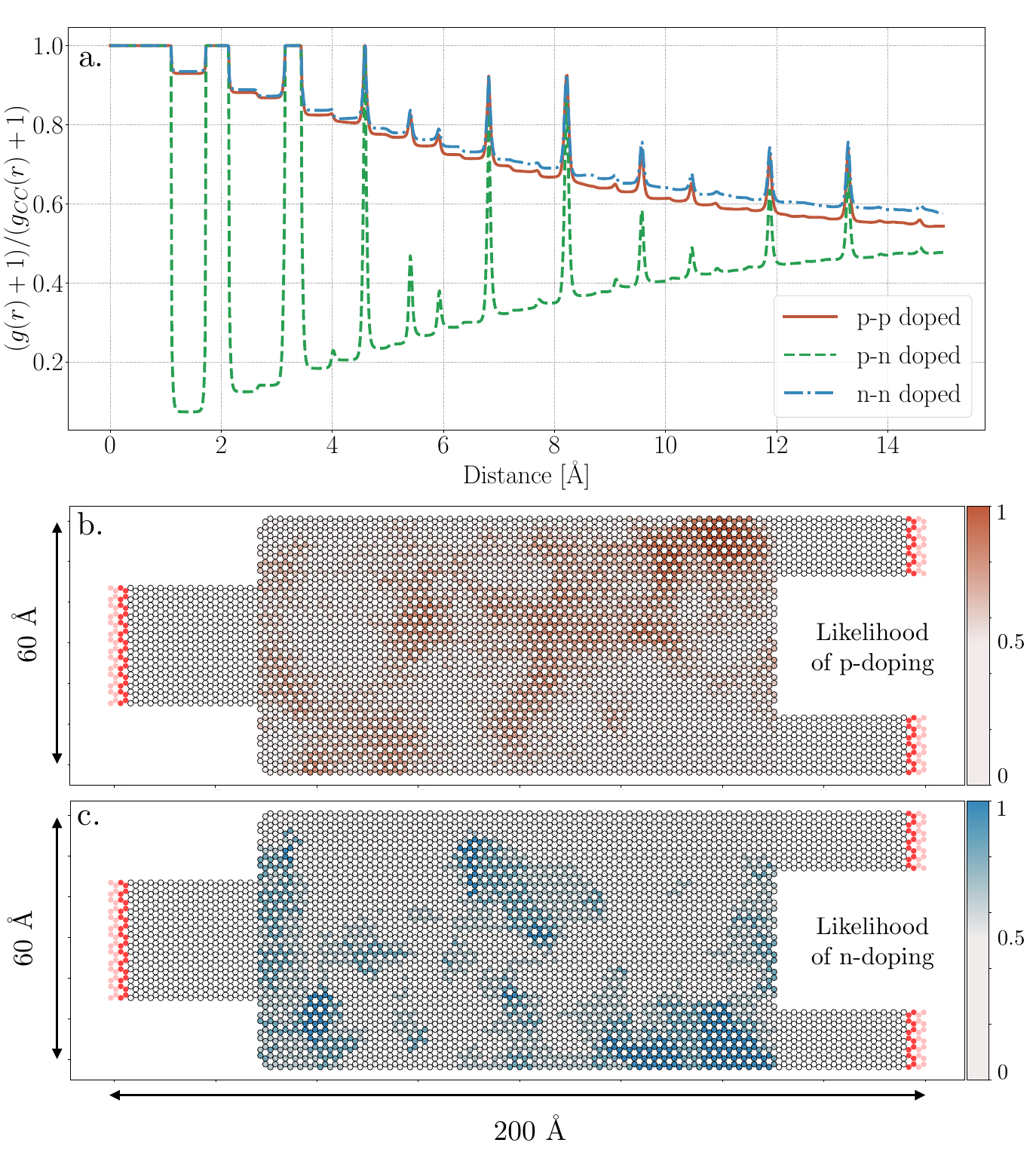}
    \caption{Pair correlation functions (PCFs - $g(r) + 1$) divided by the PCF of graphene ($g_{\text{CC}}(r) + 1$) where p- and n-doping are considered to be different atoms (a). Likelihood of finding a given lattice site with p-doping (b) and likelihood of finding a given lattice site with n-doping (c) for the optimizations where sites were either p(n)-doped or undoped. }
    \label{fig:results_2}
\end{figure}

To further understand the devices resulting from the optimization procedures, we investigate the doping profiles and their effects on valley currents. In particular, we perform similar optimizations as discussed in the Methods section but only allow for either p- or n-doping, rather than both. In Figure \ref{fig:results_2} b-c we plot the likelihood that a given site would have p- or n-doping when the valley currents are optimized. The likelihood comes from averaging over the final structures from the 64 optimizations with different seeds. We see the algorithm prefers uniform doping directly in contact with the leads, with the p(n)-doping acting as a waveguide for the $K'$($K$) valley current towards lead $1$ ($2$). Each choice of doping guides one of the valley currents and has little effect on the other valley current. In the case of p(n)-doping, the $K$($K'$) valley current had a valley current purity of $\simeq57\%$. In addition to these experiments, we also performed calculations to optimize either $K$ or $K'$ purity, allowing for both p- and n-doping, and following the same protocol as described in the Methods section. We found that we could reach 96\% purity for $K'$ and 97\% purity for $K$, which is on-par with the valley purity reported previously \cite{Rycerz2007ValleyGraphene}.\\

In contrast with this doping-induced behavior for single valley polarization, devices optimized using Eq. \ref{optimization_function} shows mixed doping in front of lead $0$, and weak long-range order, as shown in Fig \ref{fig:results_2}. To quantify to length scale of the ordering of the devices found by optimizing Equation \ref{optimization_function}, we calculated pair correlation functions (PCFs) $g(r)$ where the atom types are labelled by their respective doping. We applied Gaussian blurring with a standard deviation of 0.075~\AA~ to the PCFs to eradicate the peaks from the regular lattice structure. If $(g(r) + 1)/(g_{\text{CC}}(r) + 1)\approx1$, then we expect to see uniform doping across the crystal. If $(g(r) + 1)/(g_{\text{CC}}(r) + 1)\approx0$, then we expect to see non-uniform doping across the crystal. For $(g_{\text{pp}}(r) + 1)/(g_{\text{CC}}(r) + 1)$ and $(g_{\text{nn}}(r) + 1)/(g_{\text{CC}}(r) + 1)$, we find that the values decrease from 1 as $r$ increases. Therefore if we have a site that is p(n)-doped, we expect adjacent sites to be p(n)-doped for up to 10 \AA~ with high probability. Beyond 10 \AA, there is a $\sim50\%$ chance to see a p(n)-doped site given p(n)-doped site is selected, indicating seemingly random arrangement. The slight discrepancy between $g_{pp}(r)$ and $g_{nn}(r)$ beyond 10~\AA~in Figure \ref{fig:results_2}a is due to more sites being n-doped.

We note that this $10$ \AA~ length scale exceeds the information from neighbors and next-nearest neighbors given to the superoperator, and are a result of the ANN policy network and blurring protocol; the local environment encoded in the input vectors overlap allowing information to be propagated. Without the blurring protocol this length scale drops to 6~\AA~but still exceeds the distance between a site and it's next-nearest neighbour. The same effect can be seen when machine learning potential energy surfaces using fingerprint functions to describe atomic environments \cite{behler2007generalized, schutt2018schnet}. These fingerprint functions have a cut-off radius that may be smaller than a certain interaction distance (i.e. vdW interactions), but because of the overlap of the fingerprints, they still can describe interactions that exceed the cut-off radius.

 \begin{figure}[H]
    \centering
    \includegraphics[width=0.99\linewidth]{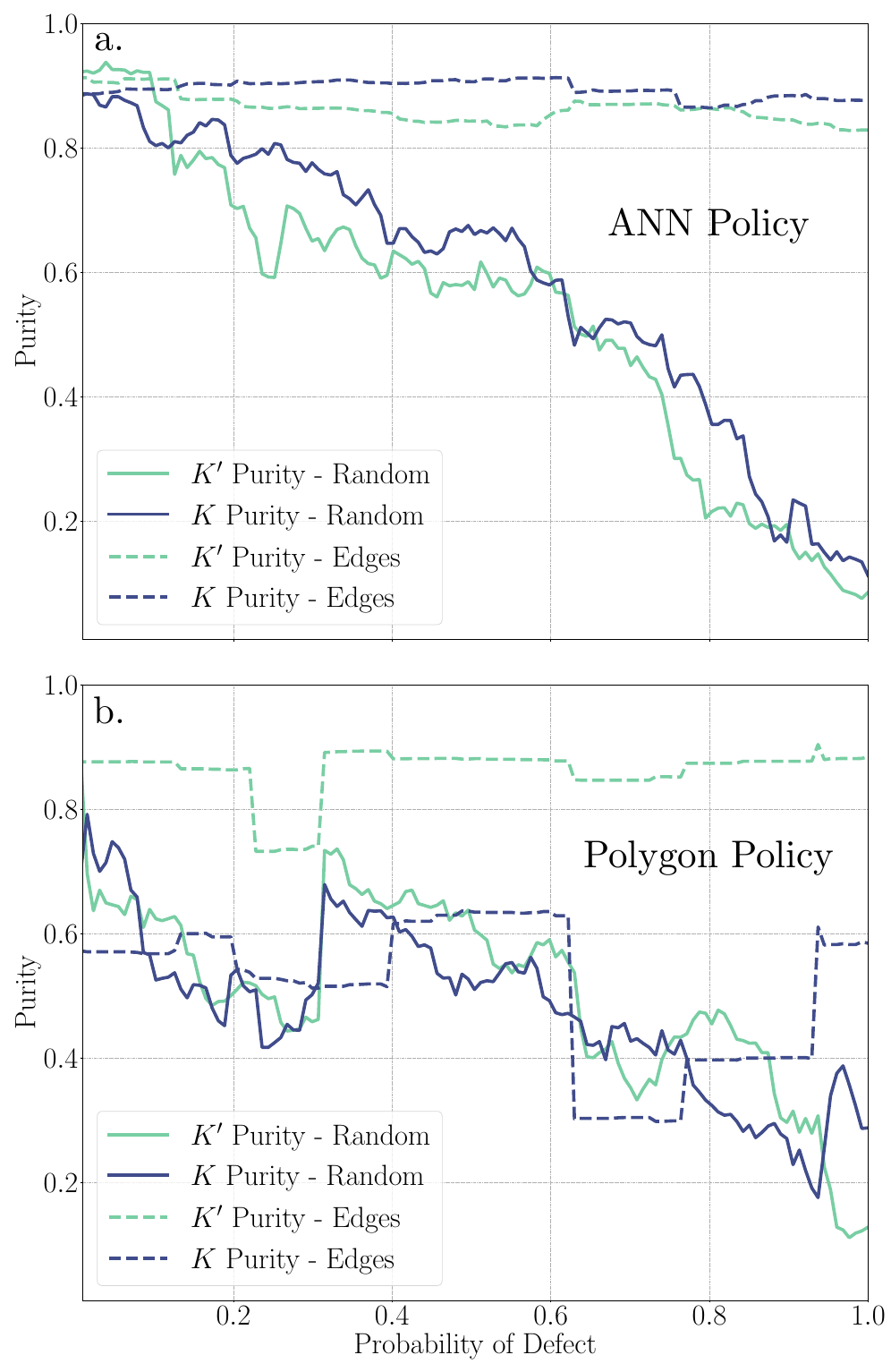}
    \caption{Purity of the most optimal device as a function of the probability that a lattice site will have a defect (flipped doping). Solid curves are for randomly selected sites and dashed curves are for sites separating p-doped regions from n-doped regions (edge sites). The top plot is for the ANN policy (a), and the bottom plot is for the polygon policy (b). See the Methods section for more information on the ANN policy and the supplemental information \cite{SI} for more information on the Polygon policy.}
    \label{fig:results_3}
\end{figure}

Lastly, we investigate the sensitivity of the generated structure that had the maximum objective function. To do so, we consider two protocols. In the first protocol that we refer to as the `random' protocol, we scan through every lattice site and flip the doping (p- to n-doping or vice versa) of the site if $u < p$ where $u$ is a randomly generated number $u\in[0,1)$ and $p$ is the probability that the doping will be flipped. In the second protocol, we only consider flipping the doping of lattice sites that separate p-doped regions from n-doped regions. We refer to these lattice sites as `edges' and refer to this protocol as the `edge' protocol. One should note that the number of edge sites makes up a small fraction of the total number of sites in the device. 

In Figure \ref{fig:results_3}, we show the purity of the valley currents as a function of the probability $p$ for the ANN policy outlined in the Methods section as well as our polygon policy outlined in the supplemental information \cite{SI}. For the ANN policy with the random protocol, we find that the purity of both $K$ and $K'$ valley currents decrease linearly (with noise) as a function of the probability $p$. This is in agreement with \cite{Rycerz2007ValleyGraphene}, where random vacancies were introduced in the lattice. For the ANN policy with the edge protocol, we find that the purity remains almost constant, with a slight decay as the probability $p$ increases. This indicates that the proposed structure is robust to changes around the edges of p- or n-doped regions. 

In contrast, for the polygon policy with the random protocol we find a decrease of the purity of valley currents as a function of the probability $p$, but with large steps at certain values of $p$. These large jumps in purity are also observed for the polygon policy with the edge protocol. These large and random jumps in the purity indicate the sensitivity of this protocol. Electron waves can be focused and split, similar to light waves, depending on the shape of the doped regions \cite{Cheianov2007TheJunctions, garcia2008fully}. When one changes the curvature of a lens slightly, the behavior of light can be drastically different. A similar process is occurring here with the electron waves.

\section{Conclusion}
In conclusion, we describe a technique that uses genetic algorithms and artificial neural network policies to optimize the purity of valley currents in graphene nanodevices with pn-doping. This optimization strategy operates on a tight-binding Hamiltonian and yields a new, optimized Hamiltonian for our objective function. This technique allows for rapid convergence of the optimization parameter studied, and yields similar solutions from independent calculations with different seeds. After averaging over an ensemble of optimization procedures we have found that p(n)-doping acts as a waveguide for $K'$($K$) valley current, allowing one to physically separate valley currents in graphene nanoribbons. After averaging over the ensemble of optimization procedures with both p- and n-doping, we found that the purity of the valley currents were $\simeq93\%$. The best-performing device allowed for a terminal-specific separation of valley currents with $\simeq 96$\% ($\simeq 94\%)$ $K$ ($K'$) valley purity. When averaging over the ensemble of optimization procedures with only p(n)-doping, we found that the purity of $K'$($K$) valley remains at $\simeq 93\%$. This shows that p(n)-doping acts as a guide to $K'$($K$) valley current. We also achieve a valley purity of 96\% for $K'$ and 97\% for $K$ current when only optimizing one valley. Additionally, we found that the artificial neural network policy can produce structures with long-range order despite only having local information. We also performed sensitivity analysis which showed that the proposed optimal structure of the artificial neural network policy is robust to edge defects. Such a device could be used to convert a quantum state to a digital signal. 
\label{conclusion}

\begin{suppinfo}
Outline and results of the optimization procedure with the polygon policy. In addition the code used for this project can be found at \url{http://clean.energyscience.ca/codes}.
\end{suppinfo}

\begin{acknowledgement}
KR and IT  acknowledge funding from the Natural Sciences and Engineering Research Council of Canada. Work at the National Research Council was carried out under the auspices of the AI4Design Program. KR and IT acknowledge Compute Canada and the National Energy Research Scientific Computing Center for computational resources. This material is based upon work supported by Laboratory Directed Research and Development (LDRD) funding from Argonne National Laboratory, provided by the Director, Office of Science, of the U.S. Department of Energy under Contract No. DE-AC02-06CH11357. Use of the Center for Nanoscale Materials, an Office of Science user facility, was supported by the U.S. Department of Energy, Office of Science, Office of Basic Energy Sciences, under Contract No. DE-AC02-06CH11357.
\end{acknowledgement}
\bibliography{refs}

\end{document}